\def   \ni {\noindent}

\def   \ssk {\vskip  5truept}

\def   \bsk {\vskip 15truept}

\def   \newline {\hfil\break}

\def\Kbs{{\,{\rm ~Kb} {\,s^{-1}\,}}}
\def\eea{\end{eqnarray}}
\def\ds{\displaystyle}

\def\ssz{\scriptsize}

\def\ni{\noindent}

\def\ln{\mbox{ln}}

\def\barc{\begin{array}{c}}
\def\ear{\end{array}}

\def\bit{\begin{itemize}}
\def\eit{\end{itemize}}

\def\and  {\it {et al.} \rm}

\def\etal{{\rm et~al. }}
\def\spose#1{\hbox to 0pt{#1\hss}}
\def\simlt{\mathrel{\spose{\lower 3pt\hbox{$\mathchar"218$}}
     \raise 2.0pt\hbox{$\mathchar"13C$}}}
\def\simgt{\mathrel{\spose{\lower 3pt\hbox{$\mathchar"218$}}
     \raise 2.0pt\hbox{$\mathchar"13E$}}}
\def\beq{\begin{equation}}
\def\eeq{\end{equation}}
\def\bce{\begin{center}}
\def\ece{\end{center}}
\def\bea{\begin{eqnarray}}
\def\eea{\end{eqnarray}}
\def\ben{\begin{enumerate}}
\def\een{\end{enumerate}}

\def\ni{\noindent}

\def\brr{\begin{array}}
\def\err{\end{array}}

\documentstyle{article}
\begin{document}
%
\def\la{\mathrel{\mathchoice {\vcenter{\offinterlineskip\halign{\hfil
$\displaystyle##$\hfil\cr<\cr\sim\cr}}}
{\vcenter{\offinterlineskip\halign{\hfil$\textstyle##$\hfil\cr
<\cr\sim\cr}}}
{\vcenter{\offinterlineskip\halign{\hfil$\scriptstyle##$\hfil\cr
<\cr\sim\cr}}}
{\vcenter{\offinterlineskip\halign{\hfil$\scriptscriptstyle##$\hfil\cr
<\cr\sim\cr}}}}}
\def\ga{\mathrel{\mathchoice {\vcenter{\offinterlineskip\halign{\hfil
$\displaystyle##$\hfil\cr>\cr\sim\cr}}}
{\vcenter{\offinterlineskip\halign{\hfil$\textstyle##$\hfil\cr
>\cr\sim\cr}}}
{\vcenter{\offinterlineskip\halign{\hfil$\scriptstyle##$\hfil\cr
>\cr\sim\cr}}}
{\vcenter{\offinterlineskip\halign{\hfil$\scriptscriptstyle##$\hfil\cr
>\cr\sim\cr}}}}}
\def\degr{\hbox{$^\circ$}}
\def\arcmin{\hbox{$^\prime$}}
\def\arcsec{\hbox{$^{\prime\prime}$}}

\hsize 5truein
\vsize 8truein
\font\abstract=cmr8
\font\keywords=cmr8
\font\caption=cmr8
\font\references=cmr8
\font\text=cmr10
\font\affiliation=cmssi10
\font\author=cmss10
\font\mc=cmss8
\font\title=cmssbx10 scaled\magstep2
\font\alcit=cmti7 scaled\magstephalf
\font\alcin=cmr6 
\font\ita=cmti8
\font\mma=cmr8
\def\ref{\par\noindent\hangindent 15pt}
\null


\title{\ni Data compression on board 
the PLANCK Satellite Low Frequency Instrument: 
optimal compression rate}                                               

\bsk 
\bsk
\author{\ni E. Gazta\~{n}aga, J. Barriga, A. Romeo, P. Fosalba, E. Elizalde}
\bsk
\affiliation{Consejo Superior de Investigaciones Cient\'{\i}ficas (CSIC) \\
\indent Institut d'Estudis Espacials de Catalunya,  \\ 
\indent Edf. Nexus-201 - c/ Gran Capit\`a 2-4, 08034 Barcelona, Spain}
\bsk
\baselineskip = 12pt

\abstract{{\bf{\scriptsize{ABSTRACT}}}
\ni Data on board the future PLANCK
Low Frequency Instrument (LFI), to measure the Cosmic Microwave Background (CMB)  anisotropies,
consist of $N$ differential temperature measurements, expanding a range
of values we shall call $R$. Preliminary studies and telemetry allocation 
indicate the need of compressing these data
by a ratio of  $c_r \simgt 10$.
Here we present a study of entropy 
for (correlated multi-Gaussian discrete)  noise,
showing how the optimal compression $c_{r,opt}$, 
for a linearly discretized data set with
$N_{bits}=\log_2{N_{max}}$ bits is given by:
$c_r \simeq {N_{bits}/
\log_2(\sqrt{2\pi e} ~\sigma_e/\Delta)}$,
where $\sigma_e\equiv (det C)^{1/2N}$ is some effective noise rms 
given by the covariance matrix $C$
and $\Delta \equiv R / N_{max}$ is the digital resolution. This
$\Delta$ only needs to be as small as the instrumental white noise  
RMS: $\Delta \simeq \sigma_T \simeq 2 mK$
(the nominal $\mu K$ pixel 
sensitivity will only be achieved after averaging).
Within the currently proposed $N_{bits}=16$ representation, a linear
analogue to digital converter (ADC) will allow the digital storage of 
a large dynamic range of differential temperature $R= N_{max} \Delta $ 
accounting for 
 possible instrument drifts and instabilities
(which could be reduced by proper on-board calibration).
A well calibrated signal will be dominated by thermal (white) noise
in the instrument: $\sigma_e \simeq \sigma_T$, which could
yield large compression rates $c_{r,opt} \simeq 8$. This is
the maximum lossless compression possible.
In practice, point sources and $1/f$ noise will produce $\sigma_e > \sigma_T$
and $c_{r,opt} < 8$.
This strategy seems safer 
than non-linear ADC or data reduction schemes (which could
also be used at some stage).
}                                                    
\ssk
\baselineskip = 12pt

\bsk
\baselineskip = 12pt


\text{\ni {\bf{1. INTRODUCTION}}
\ssk
\ni     A compression rate of about $c_r \simeq 10$ is required on board
the PLANCK Satellite LFI
(see \S 2.1 below). The data rate could be reduced by accounting
 for the relative significance of different bits (large and small
temperature differences) in the analogue-to-digital 
converters ADC 
(Herreros \etal 1997). A further compression
is assumed to be possible with classical lossless data compression 
techniques. 

Typically, 
standard lossless data compression techniques are applied 
successfully only to data sets with some redundancy.
This redundancy can be formally expressed using the entropy per
component (Shannon's entropy), $h$.
A discretized data set can be represented by $N_{bits}$,
which for a linear ADC
is typically given by the maximum range $N_{max}$:
$N_{bits}=\log_2{N_{max}}$.
If we express the joint probability for a set of N measurements
as $p_{i_1,\dots, i_N}$,
we have that the Shannon entropy per 
component of the data set is:

\beq
h \equiv - {1\over{N}} \sum_{i_1,\dots, i_N} p_{i_1,\dots, i_N}
\log_2(p_{i_1,\dots, i_N}).
\label{h}
\eeq
Shannon's theorem states that $h$ is a lower bound to the average length of
the code units.
We will define the  theoretical (optimal) compression rate as
\beq
c_{r,opt} \equiv {N_{bits}\over{h}}
\label{cr}
\eeq
For a uniform
distribution of $N$ measurements we have $p_i=1/N$ and $h= log_2 N$,
which equals the number of bits per data. Thus: {\it it is not possible to
compress a (uniformly) random distribution of  measurements}.
If noise is discretized to a high resolution 
(as compared to its variance) the resulting distribution of numbers 
approaches a uniform distribution and it is therefore virtually impossible
to compress.
This indicates that, to first approximation, it seems 
difficult to produce a lossless algorithm for compression when
the data is dominated by noise, but, as we shall see, the problem depends
crucially on the digital resolution and the range of values to
be stored.


\bsk
\ni {\bf{2. THE COMPRESSION PROBLEM}}
\ssk
\ni 
 
\ni
{\bf{2.1 Data Rate, Telemetry and compression}}
\ssk
\ni \label{dr}

\begin{table*}

\begin{center}

\begin{tabular}{|l||l|l|l|l|l|l|}
\hline
$\nu$(\scriptsize {GHz}) & \scriptsize{FWHM} & $\sigma_T$($t_s$) & Range $\Delta T$ & Det. & Rate(2.5')
& Rate($\frac{FWHM}{2.5}$) \\ \hline \hline
30 & 33' & 2.8 mK & -30-61 mK & 4 & $9.3 \Kbs$ & $1.8 \Kbs$\\ \hline 
44 & 23' & 3.2 mK & -30-138 mK & 6 & $13.9 \Kbs$ & $3.8 \Kbs$ \\ \hline
70 & 14' & 4.1 mK & -20-340 mK & 12 & $27.8 \Kbs$ & $12.4 \Kbs$\\ \hline
100 & 10' & 5.1 mK & -10-667 mK & 34 & $78.8 \Kbs$ & $49.3 \Kbs$ \\ \hline
\hline
\scriptsize {TOTAL} &  &  & -30-667 mK & 56 & $130 \Kbs$  & $67 \Kbs$\\ \hline \hline
\scriptsize {+LOAD} &  &  &   & 112 & $260 \Kbs$ &  \\ \hline 
\end{tabular}
\end{center}
\caption{{\bf{\scriptsize{Table 1.}}} Parameters for the radiometers: a) central frequency
(bandwidth is 20\%); b) angular resolution (beam FWHM); c) the RMS thermal
noise expected at 6.9 ms (144.9 Hz) sampling; d) range of temperatures 
expected from the sky (Jupiter, dipole, S-Z); e) number of detectors
(2x horns); f) total data rate at 6.9 ms (2.5 arcmin); g)
data rate for pixels of length FWHM/2.5 along the scanning circle.}
\label{radio}
\end{table*}

Following the PLANCK LFI Scientific and Technical Plan 
(Part I, \S 6.3, Mandolesi \etal 1998)
the raw data rate of the LFI is $r_d \simeq 260 \Kbs$. This assumes: 
i) a sample frequency of 6.9 ms or $f_{sampl}=144.9$ Hz, 
which corresponds to 2.5 arcmin in the sky,
1/4 of the FWHM at 100 GHz,
ii) $N_{detec}=112$ detectors: sky and reference load temperature for 56
radiometers. 
iii) $N_{bits}=16$ bits data representation.
Thus that the raw data rate is:
\beq
r_d = f_{sampl} \times  N_{detec} \times N_{bits} \simeq 259.7 \Kbs.
\eeq
The values for each channel are shown in Table 1. 
A factor of two reduction can be obtained by only transmitting the 
difference between sky and reference temperature. 
To allow for the recovery of 
diagnostic information on the separate stability of the 
amplifiers and loads,
the full sky and reference channels of a single radiometer
could be sent at a time (changing the selected radiometer from time to time
to cover all channels). 

Note that the sampling resolution of 6.9 ms corresponds to 2.5
arcmin in the sky, which is smaller than the
nominal FWHM resolution. 
Adjacent pixels in a circle could be averaged on-board to obtain the nominal 
resolution (along the circle direction). In this case the pixel size should
still 
be at least $\simeq 2.5$ smaller that the FWHM to allow for a proper
 map reconstruction. 
Note that each circle in the sky will be separated 
by 2.5' so even after this averaging along the circle scan
 there is still a lot of redundancy 
across circles. For
pixels of size $\theta \simeq FWHM/2.5$ along the circle scan
the total scientific rate could be reduced to $r\simeq 67 \Kbs$
as shown in Table 1 (or $134 \Kbs$ with some subset information of the ref. load).

\ni The telemetry allocation for the
LFI scientific data is expected to be $r_t=20 \Kbs$.
Thus the target compression rates
are about:

\beq
c_r = {r_d\over{r_t}} \simeq 3- 13,
\eeq
depending on the actual on-board processing and
requirements.

\bsk
\ni {\bf{2.2 Data Structure}} 
\ssk
\ni Planck's satellite spins with a frequency $f_{spin}=1$ rpm
so that the telescope (pointing at approximately right angle to the spin axis)
sweep out nearly great circles in the sky. Each circle is scanned  
over 1 (or 2) hours, so that there are 
60 (or 120) images of the same pixel. Each measurement is mostly 
dominated by instrumental noise, 
$\sigma_T \sim 2 mK$ (see Table 1)
rather than by the CMB noise ($\sigma_{CMB} \simeq 10^{-2} mK$).
If this noise (at frequencies smaller
than $f_{spin}$) is mostly 
thermal, there is no redundancy in these images and 
little hope for compression. But one could then say that
in this case there is no need for compression, 
as we can just average those 
60 images of a pixel
and only send the mean downwards to Earth. 
The problem is that 
one expects $1/f$ noise to dominate the
instrument noise at frequencies smaller than $\sim 0.1$ Hz. 
Thus, compression is only required when we want to keep these 60 
(120) images
in order to correct for the instrument instability in the data 
reduction process (on Earth). This $1/f$ type of
noise is more redundant and might be subject to some compression,
but even in this case if we keep it to a high resolution 
(as compared to its $rms$) the resulting  probabilities
would be close to a uniform distribution and compression would be
nearly impossible.

\bsk
\ni {\bf{2.3 Dynamic range \& calibration}}
\ssk
\ni The final dynamic range for the measured temperature differences
per angular resolution pixel will be $\Delta T \simeq 1 \mu K- 1K$. The 
lowest resolution
of $\simeq 1 \mu K$ will only be obtained after averaging all data.
The highest $\simeq 1K$ being the hottest source that we want
to keep (not saturated) by anyone of the frequencies.
Positive signals from Jupiter, which will be used for calibration,
can be as large as $\simeq 0.7K$ at 100 Ghz. Other point sources and
the Galaxy give intermediate positive values. 
Negative differences (with respect to the 
mean CMB $T \simeq 2.7K$), of the order of a few $mK$, can be originated 
by the dipole, the relative  velocity  between the
satellite velocity and the CMB rest frame. The Sunyaev-Zeldovich effect
(towards a total of a few hundreds Clusters of Galaxies) 
can also give a negative signal of few $10 mK$. Thus the overall range of
mesurements is $-30 mK$ to $1 K$.

As pointed out by Herreros \etal (1997)
the temperature resolution
is given by the receiver noise $\sigma_T$ on the sampling
time 6.9 ms (or corresponding value if there is some on-board averaging)
and not by the final target sensitivity.
At the end of the mission, each FWHM pixel will have been measured 
$\simeq 10^6$ times. Thus a lower resolution $\Delta T \simeq 1 \mu K$
is not necessary on board, given that the raw signal is dominated 
by the white noise component. This higher resolution will be later obtained
by the pixel averaging (data reduction on Earth). 

We can distinguish two basic components for the receiver noise: the white
or thermal noise, and the instabilities or calibration gains
(like the $1/f$ noise). An example is given by 
the following power spectrum of 
frequencies $f$:

\beq
P(f) = A \left(1 + {f_{knee}\over{|f|}}\right).
\label{pk}
\eeq
The 'knee' frequency, $f_{knee}$, is expected to be $f_{knee} \simeq 0.005$ Hz
for a 4K load or  $f_{knee} \simeq 0.06$ Hz
for a 20K load.
The expected RMS thermal noise, $\sigma_T \propto A$
at the sampling frequency (2.5 armin), 
is listed in Table 1. The lowest value is
given by the 30 Ghz channel and could be further reduced to $\simeq 1 mK$
if the data is averaged to FWHM/2.5 to obtain the nominal resolution.
The larger values in the dynamical range can be affected by the calibration 
gains. This is important and should be carefully taken into account if
a non-linear ADC is used, as gains could then change the relative significance
of measurements (eg, less significant bits shifting because of gains).
In fact, a $1/f$ power spectrum integrated from the knee-frequency 
($f_{knee}$ for a time $T$, gives a diverging rms noise:

\beq
\sigma_{1\/f}^2= {\sigma_T^2\over{f_{max}}} \int_{1/T}^{f_{max}} 
df \, {f_{knee}\over{f}} 
= \sigma_T^2 {f_{knee}\over{f_{max}}} \, \ln{(T \, f_{max}})
\eeq

For a $T \simeq 1$ year mission the contribution from the $1/f$ noise
in pixels averaged after
succesive pointings $f_{max} \simeq 10^{-4}$ and we have
 $\sigma_{1\/f}^2 \simeq 10^4 \sigma_T^2$! This illustrates
why the calibration problem is so important and makes a large
dynamic range desiderable. Averaging pixels at the
spin rate, $f_{max} \simeq f_{spin}$, gives 
$\sigma_{1\/f}^2 \simeq 10 \sigma_T^2$, this is not
too bad for the dynamic range, but it corresponds to a mean value and
there could be more important instantaneous or temporal gains.
Drifts with periods longer than the spin period (1 rpm) can be removed by requiring 
that the average signal over each rotation at the same pointing remains constant.
Drifts between pointings (after 1 or 2 hours) could be reduced by using the 
overlaping pixels. All this can be easily done on-board, while a more careful
matching is still possible (and necessary) on Earth. This allows the on-board gain
to be calibrated on timescale larger than 1 min with an accuracy given by 
$\sigma_T$. Additional and more carefull in-flight
calibration can also be done using the the signal from external planets
and the CMB dipole. In any case we will assume here that instabilities
or gains are under control ($\simeq \sigma_T$) for frequencies larger than the 
spin frequency. For smaller times we will typically use Eq.(5)
as a mean value but bearing in mind that larger gains are also possible.

In summary, because of the possible instrument gains,
 it is impotant to have a constant resolution of $\simeq \sigma_T \simeq 1 mK$
over a large range of values ($\Delta T \simeq 1K$) to be able to
recover the underlaying signal after proper calibration. This could
be partially done on board. A constant resolution
indicates the need of a linear ADC,
which with adequate compression (presented next) will be shown to be 
a good alternative to non-linear ADC.
\bsk
\ni {\bf{3. A SOLUTION TO THE PROBLEM}} 
\ssk
\ni In a separate paper (Romeo \etal  1998) we have presented
a general study of (correlated multi-Gaussian) noise compression 
by studying  Shannon entropies per componet $h$, and therefore the
optimal compression $c_{r,opt}$ in Eq.(2).
For a linearly discretized data with
$N_{bits}=\log_2{N_{max}}$ bits, $h$ in Eq.(1)
depends only on the ratio of the digital resolution 
$\Delta$  to the
effective rms noise, $\sigma_e$: 
\beq
h  = \log_2(\sqrt{2\pi e} ~\sigma_e/\Delta)
\eeq
with $\sigma_e^2 \equiv (det C)^{1/N}$, 
where $C$ is the covariance
matrix of the (multi-Gaussian random) field $x_i$:
$C_{ij} \equiv <x_i x_j>$. 
As mentioned in the section above, $\Delta$ in Planck
only needs to be as small as the instrumental white noise  $\sigma_T$.
If the data is dominated by thermal instrumental noise we have
$\sigma_e \simeq \sigma_T \simeq \Delta$ and the optimal compression
is simply: 

\beq
c_{r,max}=N_{bits}/\log_2(\sqrt{2\pi e}) \simeq 8.
\eeq
where we have use $N_{bits}=16$ as planned for the Planck LFI.
This is the {\it  maximum} lossless compression that can be achieved
for a well calibrated signal dominated by  instrumental
thermal (white) noise.
This very large compression rate can be obtained because there is a large
range of values $\simeq \Delta 2^{N_{bits}}$  which
 has a very small probability, and therefore can be easily compressed
(e.g. by Huffman's or arithmetic coding)
but can't be omitted because they are needed to calibrate the
instrument gains (and to  measure the  point sources and the galaxy). 
In practice point sources and $1/f$ noise will produce $\sigma_e > \sigma_T$
and $c_{r,opt} < 8$ (note that in general $\sigma_e< \sigma_0$, the one-point
RMS fluctuation) and will also need to account for the fact that the
galaxy and the point sources can't be represented by a multi-Gaussian,
even if we allow for a different power spectrum.
Lets do a more realistic case with 1/f noise, but still without point sources
or the galaxy.

\begin{table}

\begin{center}

\begin{tabular}{|l|l|l|l|l|l|}
\hline
 $f_{knee}$ & $f_{min}$ & $f_{max}$ & $f_L (Hz)$ & $A'/A$ & $C_{r,opt}$ \\
\hline
0 & 0.017 & 145     & 0 & 1 & 7.816 \\
0.005 & 0.017 & 145 & 0 & 1 &7.814 \\
0.06 & 0.017 & 145  & 0& 1 & 7.798 \\
0.06 & 0.017 & 10  & 0& 1 & 7.754 \\
1.00 & 0.017 & 145  & 0& 1 & 7.750 \\
0.005 & 0 & 145  &  0.017& 1 & 7.814 \\
0.06 & 0 & 145  &  0.017& 1 & 7.799 \\
0.06 & 0 & 145  &  0.017& 10 & 7.798 \\
\\\hline
\end{tabular}

\end{center}
\caption{{\bf{\scriptsize{Table 2.}}} Compression factors for different instrumental noise.}
\label{rates}
\end{table}

\bsk
\ni {\bf{3.1 Compression with 1/f noise}}
\ssk
\ni 

In the case of the power spectrum in Eq.(5), Romeo \etal (1998)
find for the entropy:
\beq
h-h_0 \approx {1\over{2}} \log _2\left[1+\frac{f_{knee}}{f_{max}}\right]+
{1 \over 2}\frac{f_{knee}}{f_{max}}\log _2\left[\frac{f_{max}+f_{knee}}{f_{min}+f_{knee}}\right]
\eeq
where $h_0$ is the white noise (thermal) contribution [$h_0 
= \log_2(\sqrt{2\pi e} ~\sigma_T/\Delta)$] and $f_{max}$ and $f_{min}$
are the maximum and minimum frequencies cover with the N measurements.
In our case we are only interested in the contribution within a revolution
(as we are assuming a good calibration at smaller frequencies) so that
 $f_{max}= f_{sampl} \simeq 145$ Hz and $f_{min} \simeq f_{spin} \simeq 1/60$ Hz.

Another case which can be of interest is:
\beq
P(f)=\left\{  \begin{array}{ll}
A', & \mbox{for $f \le f_{L}$,} \\
\ds A+{f_{knee}\over|f|} ,
& \mbox{for $f_{L} < f \le f_{\mbox{\ssz max}}.$}
\end{array}
\right.
\eeq
Taking again as reference $h_0$ as the thermal 
 case where $f_{knee}=0$ and $A'=A$, we may
write
\begin{eqnarray}
h&\simeq&h_0+\frac{1}{2}
\log_2\left[ 1+\frac{f_{knee}}{f_{max}} \right]
-\frac{1}{2} \frac{f_L}{f_{max}}
\log_2\left[ 1+\frac{f_{knee}}{f_{L}} \right]
+
\nonumber\\
&&+\frac{1}{2}\frac{f_{knee}}{f_{max}}
\log_2\left[\frac{ 
f_{max}+f_{knee}}{f_{L}+f_{knee}}\right]
+\frac{1}{2}\frac{f_L}{f_{max}}
\log_2\left[\frac{A'}{A }\right]
\end{eqnarray}

In Table 2
we have presented some compression rate values
corresponding to this entropy for different 'knee' frequencies.

\ni In order to quote a more realistic compression rate, it 
is crucial to have a detailed model
of the instrument instabilities (i.e., what is the value of $f_{knee}$ frequency?
what is the value of the white noise amplitude?), the detailed ADC model,
the on-board data and calibration strategy, the pointing 
and a detailed simulations of the sky.


\bsk
\ni {\bf{5. CONCLUSION}}
\ssk
\ni Because of the possible instrument gains,
 it is impotant to have a constant resolution of $\simeq \sigma_T \simeq 1 mK$
over a large range of values ($\Delta T \simeq 1K$).
This indicates the convenience of a linear ADC. Although some 
compression can be achieved with non-linear ADC, in this case
standard linear lossless data compression techniques
seem safer (because of possible calibration drifts) 
and more efficient (because of the larger compression rates).

The  {\it  maximum} lossless compression that can be achieved
with a well calibrated signal is $c_r \simeq 8$ (with data of
 $N_{bits}=16$). Similar values can be obtained (\S 3.1) 
even for non-thermal instrumental noise. 
These results  assume that the dominant component of data is
multi-Gaussian (correlated) noise. Although this might be true for 
the mean instrumental noise and the CMB signal, it is not true for
point sources or the galaxy. Nevertheless, a compression factor 
of $c_r \simeq 3$
can be easily obtained if we assume that 
at least $\simeq 75\%$ of the data is drescribed  by well-calibrated
instrumental noise
(the CMB is only relevant after averaging over many pixels).

Thus, compression factors in the range $c_r \simeq 3-7$ are possible
within the approximation for the data structure we have considered.
Compression of the raw data including the reference load 
($r_d \simeq 260 \Kbs$) does not seem possible with a telemetry rate
of $r_t \simeq 20 \Kbs$, but it might be possible for $r_t \simeq 40 \Kbs$.
An alternative is to process on-board
 some of the current redundancy 
by stucking nearby pixels to the level of the nominal resolution
(FWHM/2.5), as indicated in Table 1.
The actual compression can be achieved with standard
Huffman's or arithmetic coding, although other possibilities can
also be considered or taylored for this problem (see Romeo \etal 1998 for
more details).

}
\bsk
\baselineskip = 12pt
{\abstract \ni ACKNOWLEDGMENTS This work was in part supported by Comissionat per Universitats i 
Recerca, Generalitat de Catalunya, grants ACES97-22/3 and ACES98-2/1,
by the Spanish Nacional Space Programe, CICYT, grant ESP96-2798-E and
DGES (MEC), project PB96-0925. We wish to thank the organizers of the UIMP workshop
"The CMB and the Planck Mission" in Santander for a very enjoyable meeting.}

\bsk
\baselineskip = 12pt


{\references \ni {\bf{REFERENCES}}
\ssk

\ref Herreros, J.M., Hoyland, R., Rebolo, R., Watson, R.A., 1997 Ref.:LFI-IAC-TNT-001
\ref Mandolesi, N. \etal, 1998, LFI for Planck, a proposal to ESA's AO.
\ref Romeo, A., Gazta\~naga, E., Barriga J., Elizalde, E., 1998, physics-ph/9809004 \\
 (http://xxx.unizar.es/abs/physics/9809004)

}                      

\end{document}